\documentclass[twocolumn,twocolappendix]{aastex63}

\usepackage{newtxtext,newtxmath}
\usepackage[T1]{fontenc}

\DeclareRobustCommand{\VAN}[3]{#2}
\let\VANthebibliography\thebibliography
\def\thebibliography{\DeclareRobustCommand{\VAN}[3]{##3}\VANthebibliography}

\usepackage{graphicx}	%
\usepackage{amssymb}	%
\usepackage{xspace}     %
\usepackage{xcolor}     %
\def\Mdot{\ensuremath{\dot{M}}\xspace}
\def\Mdotcr{\ensuremath{\dot{M}_{\mathrm{cross}}}\xspace}  %
\def\MP{\ensuremath{M_{\mathrm{p}}}\xspace}
\def\RP{\ensuremath{R_{\mathrm{p}}}\xspace}
\def\Rtrunc{\ensuremath{R_{\mathrm{trunc}}}\xspace}
\def\ff{\ensuremath{f_{\mathrm{fill}}}\xspace}
\def\ffill{\ff}
\def\Teff{\ensuremath{T_{\mathrm{eff}}}\xspace}
\def\nup{\ensuremath{n_{\mathrm{up}}}\xspace}  %
\def\Lya{Ly\,$\alpha$\xspace}   %
\def\Lyb{Ly\,$\beta$\xspace}    %
\def\Lyg{Ly\,$\gamma$\xspace}   %
\def\Ha{H\,$\alpha$\xspace}     %
\def\Hb{H\,$\beta$\xspace}      %
\def\Hg{H\,$\gamma$\xspace}     %
\def\Hd{H\,$\delta$\xspace}     %
\def\Paa{Pa\,$\alpha$\xspace}   %
\def\Pab{Pa\,$\beta$\xspace}    %
\def\Pag{Pa\,$\gamma$\xspace}   %
\def\Pad{Pa\,$\delta$\xspace}   %
\def\Bra{Br\,$\alpha$\xspace}   %
\def\Brb{Br\,$\beta$\xspace}    %
\def\Brg{Br\,$\gamma$\xspace}   %
\def\Brd{Br\,$\delta$\xspace}   %
\def\nuHa{\nu_{\mathrm{H}\,\alpha}} 
\def\LHa{\ensuremath{L_{\mathrm{H}\,\alpha}}\xspace}
\def\Lacc{\ensuremath{L_\mathrm{acc}}\xspace}
\def\Lline{\ensuremath{L_\mathrm{line}}\xspace}
\def\Rin{\ensuremath{R_\mathrm{in}}\xspace}
\def\RHill{\ensuremath{R_\mathrm{Hill}}\xspace}
\def\siglgLacc{\ensuremath{\sigma_{\log\Lacc}}\xspace}
\def\siglgLHa{\ensuremath{\sigma_{\log\LHa}}\xspace}
\def\sigform{\ensuremath{\sigma^{\mathrm{form}}}\xspace}
\def\Lsun{L_\odot}
\def\Msun{M_\odot}
\def\MJ{M_\mathrm{J}}
\def\RJ{R_\mathrm{J}}
\def\vf{v_\mathrm{ff}}
\def\MJyr{\MJ/\mathrm{yr}}
\def\PDSb{\mbox{PDS\,70\,b}\xspace}                %
\def\PDSc{\mbox{PDS\,70\,c}\xspace}                %
\def\Tmax{\ensuremath{T_{\mathrm{max}}}\xspace}

\newcommand{\revise}{}      
\newcommand{\srevise}{}
\newcommand{\gab}{}
\def\kms{\mathrm{km}\,\mathrm{s^{-1}}}             %

\renewcommand{\S}{Section}

\defcitealias{Alcala+2017}{Al17}
\defcitealias{Aoyama+2018}{AIT18}
\defcitealias{Aoyama+Ikoma2019}{AI19}
\defcitealias{Aoyama21}{AMMI21}
\defcitealias{Haffert+2019}{H19}
\defcitealias{Komarova+Fischer2020}{KF20}  %
\defcitealias{Rigliaco+2012}{R12}
\defcitealias{storey95}{SH95}
\defcitealias{Szulagyi+Ercolano2020}{SzE20}
\defcitealias{Thanathibodee+2019}{Th19}

\graphicspath{{./}{figures/}}

\begin{document}

\title[Comparison of Planetary H\,$\alpha$-emission models]{%
Comparison of planetary H\,$\alpha$-emission models: A new correlation with accretion luminosity
}

\correspondingauthor{Yuhiko Aoyama}
\email{yaoyama@tsinghua.edu.cn}

\author[0000-0003-0568-9225]{Yuhiko Aoyama}
\thanks{Former Visiting Scholar of the Deutsche Forschungsgemeinschaft (German\\Research Foundation; DFG) SPP~1992 program}
\affiliation{Institute for Advanced Study, Tsinghua University, Beijing 100084, People's Republic of China}
\affiliation{Department of Astronomy, Tsinghua University, Beijing 100084, People's Republic of China}
\affiliation{Department of Earth and Planetary Science, The University of Tokyo, 7-3-1 Hongo, Bunkyo-ku, Tokyo 113-0033, Japan}

\author[0000-0002-2919-7500]{Gabriel-Dominique Marleau}
\affiliation{%
Institut f\"ur Astronomie und Astrophysik,
Universit\"at T\"ubingen,
Auf der Morgenstelle 10,
72076 T\"ubingen, Germany
}
\affiliation{%
Physikalisches Institut,
Universit\"{a}t Bern,
Gesellschaftsstr.~6,
3012 Bern, Switzerland
}
\affiliation{%
Max-Planck-Institut f\"ur Astronomie,
K\"onigstuhl 17,
69117 Heidelberg, Germany
}

\author[0000-0002-5658-5971]{Masahiro Ikoma}
\affiliation{Department of Earth and Planetary Science, The University of Tokyo, 7-3-1 Hongo, Bunkyo-ku, Tokyo 113-0033, Japan}
\affiliation{Division of Science, National Astronomical Observatory of Japan, 2-21-1 Osawa, Mitaka, Tokyo 181-8588, Japan}
\affiliation{Department of Astronomical Science, The Graduate University for Advanced Studies, SOKENDAI, 2-21-1 Osawa, Mitaka, Tokyo 181-8588, Japan}

\author[0000-0002-1013-2811]{Christoph Mordasini}
\affiliation{%
Physikalisches Institut,
Universit\"{a}t Bern,
Gesellschaftsstr.~6,
3012 Bern, Switzerland
}

\begin{abstract}
Accreting planets have been detected through their hydrogen-line emission, specifically H\,$\alpha$. To interpret this, stellar-regime empirical correlations between the H\,$\alpha$ luminosity $L_\mathrm{H\,\alpha}$ and the accretion luminosity $L_\mathrm{acc}$ or accretion rate $\dot{M}$ have been extrapolated to planetary masses, however without validation.
We present a theoretical $L_\mathrm{acc}$--$L_\mathrm{H\,\alpha}$ relationship applicable to a shock at the surface of a planet.
We consider wide ranges of accretion rates and masses and use detailed spectrally-resolved, non-equilibrium models of the postshock cooling.  %
The \revise{new} relationship \revise{gives a markedly higher $L_\mathrm{acc}$ for a \revise{given} $L_\mathrm{H\,\alpha}$} than fits to young stellar objects\srevise{, because \Lya, which is not observable, carries a large fraction of \Lacc}.
Specifically, an $L_\mathrm{H\,\alpha}$ measurement needs ten to 100 times higher $L_\mathrm{acc}$ and $\dot{M}$ than previously predicted, which may explain the rarity of planetary H\,$\alpha$ detections.
We also compare the $\dot{M}$--$L_\mathrm{H\,\alpha}$ relationships coming from the planet-surface shock or implied by accretion-funnel emission. Both can contribute simultaneously to an observed H\,$\alpha$ signal but at low (high) $\dot{M}$ the planetary-surface shock (heated funnel) dominates. Only the shock produces Gaussian line wings.
Finally, we discuss accretion contexts in which different emission scenarios may apply, putting recent literature models in perspective, and also present $L_\mathrm{acc}$--$L_\mathrm{line}$ relationships for several other hydrogen lines.
\end{abstract}

\keywords{%
Extrasolar gas giants~(509);
Scaling relations~(2031);
Accretion (14);
Shocks (2086);
H~\textsc{i} line emission~(690);
H alpha photometry~(691);
Planet formation~(1241);
Classical T Tauri stars~(252) %
}

\section{Introduction}
  \label{sec:intro}
Recent observations have detected \Ha
emission from planets around young accreting stars (\citealp{Wagner+2018}; \citealp[][hereafter \citetalias{Haffert+2019}]{Haffert+2019}; \citealp{Hashimoto+2020}; \citealp{Eriksson+2020}).
For stars, sufficiently strong
\Ha 
\gab{indicates} gas accretion \citep[][]{Hartmann+2016}, \gab{and}
empirical relationships %
between \Ha luminosity and accretion luminosity exist, where the latter is estimated from UV continuum observations \citep[e.g.,][]{Fang+2009}. 
Because initially no \srevise{\LHa--\Lacc correlations} were available for the planetary case, 
these stellar scalings have been extrapolated to analyse individual detections or surveys results (\citealp{Sallum+2015,Wagner+2018}; \citetalias{Haffert+2019}; \citealp{Cugno+2019,Zurlo+2020,Xie+2020}). 
However, verifying whether these correlations hold also at planetary masses was not yet possible.

Following the reports on planetary \Ha\ detection, %
theoretical work
has attempted to reproduce and interpret the observations. 
\citet[][hereafter \citetalias{Thanathibodee+2019}]{Thanathibodee+2019} 
applied a magnetospheric accretion model developed for T~Tauri stars \citep{Muzerolle+2001} to planetary masses and radii, and could
reproduce the %
\Ha line of \PDSb.
They assumed a strong magnetic field able to truncate the accretion disc \citep{Christensen+2009,Batygin2018} and hot gas ($T\sim10^4$~K) in the accretion funnel.%

In another direction,
\citet[][hereafter \citetalias{Aoyama+2018}]{Aoyama+2018} constructed the first emission model of \textit{shock}-heated gas 
for planetary masses,
focusing on hydrogen lines. 
There, the \Ha comes from the postshock gas and not the accretion flow.
This  %
can reproduce the observations if a strong shock of preshock velocity $v\gtrsim30~\kms$ %
occurs on the circumplanetary disc (CPD) surface 
or on the planetary surface \citep[][hereafter \citetalias{Aoyama+Ikoma2019}]{Aoyama+Ikoma2019}. 
The former is suggested by isothermal 3D hydrodynamic simulations \citep{Tanigawa+2012}, in which the gas flows almost vertically in free-fall onto the CPD.
A planetary-surface shock can occur when the gas falls directly from the upper layers of 
the protoplanetary disc (PPD), 
for instance, from meridional circulation \citep{Szulagyi+2014,Teague+2019} or through magnetospheric accretion columns originating at the inner edge of the CPD \citep[e.g.,][]{Lovelace+2011,Batygin2018}. 
\revise{Such flow patterns need non-isothermal or magnetic effects, respectively.}

In this Letter, we derive new theoretical
\Lacc--\LHa and \Mdot--\LHa
relationships from our shock emission model
for planetary-surface accretion\footnote{%
   Contrary to statements  %
   in the literature, in \citetalias{Aoyama+2018} the line flux
   is not intrinsically high; it depends on the input parameters.
   Also, the $T\sim10^4$--$10^6$~K in \citetalias{Aoyama+2018}  %
   are not effective temperatures but rather part of non-equilibrium cooling in a thin postshock layer (roughly the Zel'dovich spike).}.
We compare them with
correlations measured
for stars. Afterward, we discuss the differences among theoretical models and predictions, including  %
\citet[][hereafter \citetalias{Szulagyi+Ercolano2020}]{Szulagyi+Ercolano2020}.
We also comment in Appendix~\ref{sec:Zhu} on the \LHa estimate by \citet{Zhu2015},
and present correlations for several other lines in Appendix~\ref{sec:morelines}.

\section{Stellar and planetary accretion relationships}

\subsection[Comparison of Lacc--LHa relationships]{Comparison of \Lacc--\LHa relationships}
  \label{sec:compLacc}

\begin{figure*}
\begin{center}
    \includegraphics[width=0.8\textwidth]{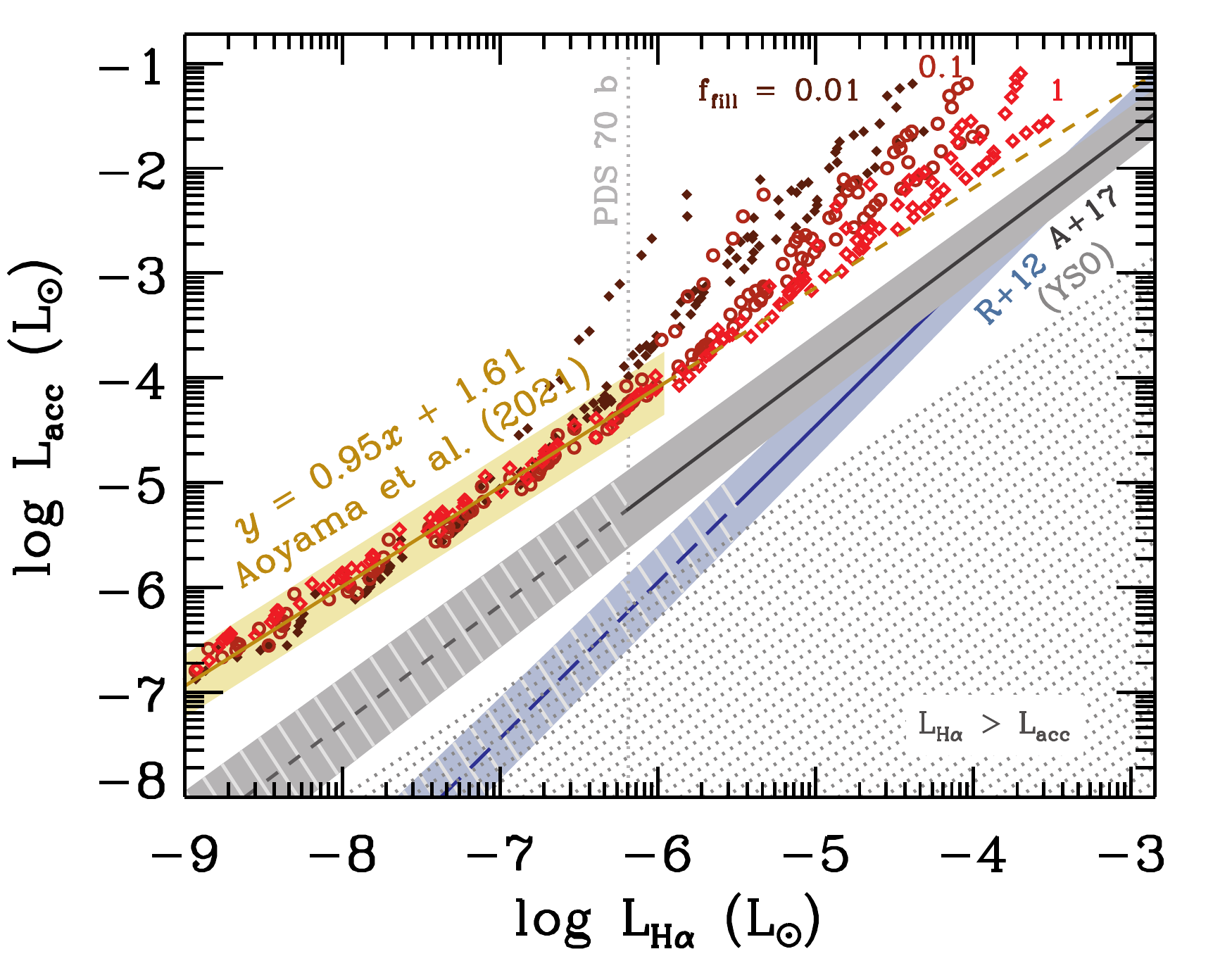}
    \caption{Theoretical relationships between accretion luminosity,
    \Lacc and \Ha luminosity, \LHa. 
    The symbols show our model results for a wide range
    of accretion rates
    $\Mdot=3\times10^{-10}$--$3\times10^{-5}~\MJyr$
    and %
    masses $\MP=2$--20~$\MJ$,  %
    with filling factor
    $\ff=0.01$ (filled diamonds), $0.1$ (circles), and $1$ (open diamonds).
    The golden line indicates
    Equation~(\ref{eq:Lacc_LHa}) %
    which fits our results 
    up to $\log(\Lacc/\Lsun)=-4$; the dashed line is an extrapolation. %
    The shaded golden region shows the spread $\pm\sigma=\pm0.3$~dex.
    Fits by \citet[][blue]{Rigliaco+2012} and \citet[][gray]{Alcala+2017} for stellar-mass objects are also shown by the blue and gray lines, respectively.
    The shaded regions reflect the formal errors (Equation~(\ref{eq:sigLL}), with $\siglgLHa=0$), which corresponds to the usual approach %
    but ignores the spread of their data ($\pm\sigma\approx\pm0.5$--0.7~dex).
    The dashed lines indicate extrapolations. %
    Extinction by material around the planet (not included) would only move the points to the left, away from the stellar relationships.
    Already without considering extinction, our %
    relationship differs clearly from the stellar fits,
    by up to %
    2.5~dex here.
    The dotted region has $\LHa>\Lacc$, which could be %
    unlikely
    (see text).  %
    }
    \label{fig:Lacc_LHa}
\end{center}
\end{figure*}

In stellar observations, UV/optical continuum measurements \citep[e.g.,][]{Gullbring+1998} have been used to estimate the accretion luminosity \Lacc by modeling the emission from
the shock-heated photosphere \citep[e.g.,][]{Calvet+Gullbring1998}. 
However, for distant objects, interstellar extinction prevents the detection of such continua. On the other hand,
\Ha is brighter and less extincted. %
Thus, empirical \Lacc--\LHa relationships derived for nearby stars are used to estimate \Lacc from the observed \LHa. Then, assuming a mass and radius or using known estimates from photometry,
\Mdot is estimated for distant accretors. %

In Figure~\ref{fig:Lacc_LHa}, we show the \Lacc--\LHa correlation from \citetalias{Aoyama+Ikoma2019}'s models, detailed in \citet[hereafter \citetalias{Aoyama21}]{Aoyama21}. 
\revise{They simulated the radiative transfer of hydrogen lines in the 1D plane-parallel flow of the shock-heated gas. Since the timescale of temperature change is comparable to that of line emission process, they numerically calculated the time-evolving electron transitions via collision and radiation. This model estimates the hydrogen line intensity for two input parameters of pre-shock gas velocity $v_0$ and number density $n_0$. Assuming the accreting gas falls onto the planetary surface with the free-fall velocity, the model estimates the hydrogen line luminosity as a function of the planetary mass $\MP$ \revise{and the} accretion rate $\Mdot$%
.}
%
As in \citetalias{Aoyama21},
$\Lacc = G \MP \Mdot (\RP^{-1}-R_\mathrm{in}^{-1})$, where $G$ is the gravitational constant, $\RP$ is the planetary radius and $\Rin$ is the radius from which the gas starts at rest.
We consider a wide range of mass accretion rates $\Mdot=3\times10^{-10}$--$3\times10^{-5}~\MJyr$ and masses $\MP=2$--20~$\MJ$, and consider a filling factor $\ff=0.01$, $0.1$, or~$1$, where the \Ha emission comes from the area \revise{of the shock of} $\ff4\pi\RP^2$. 
For $\ff=1$,
$\Rin=\infty$ (since $\Rin\approx1/4\RHill\gg\RP$, where $\RHill$ is the Hill radius; \citealp{Mordasini+2012II}), and $\Rin=5 \RP$ for $\ff\leqslant0.1$ as for magnetospheric accretion \citep{Hartmann+2016}.
We use fits by \citetalias{Aoyama21} of $\RP(\Mdot,\MP)$ from the \citet{Mordasini+2012II} planet-structure model, which predicts $\RP\approx1.5$--5~$\RJ$.

For $\Lacc\lesssim10^{-4}~\Lsun$,
\Lacc and \LHa correlate %
well with each other\footnote{Or for $\LHa\lesssim10^{-5.5}~\Lsun$ at $\ff\gtrsim0.3$
 and
$\LHa\lesssim10^{-6.5}~\Lsun$ for all \ff.}.
The spread in \Lacc ($\lesssim1.5$~dex) at higher \Lacc reflects
the large optical depth at %
\Ha. In %
high-\Lacc (or higher-density) cases, \Ha from optically thick regions hardly escapes, %
and other lines take over the energy transfer. %
Decreasing \ff increases the (pre- and) postshock density and thus the postshock optical thickness \citepalias{Aoyama+2018}.
Thus, decreasing \ff also increases \Lacc at a given \LHa, and $\ff=1$ yields the minimal \Lacc for %
a given \LHa. Also, lower masses sit towards higher \Lacc for a given \LHa because $v_0$ is lower, causing a lower excitation degree and less effective hydrogen line emission.

We fit %
our \Lacc--\LHa relationship in the form of %
$y=ax+b$
for $\Lacc\leqslant10^{-4}~\Lsun$: %
\begin{equation}
 \label{eq:Lacc_LHa}
  \log_{10}\left(\Lacc/\Lsun\right) = 0.95 \times \log_{10}\left(\LHa/\Lsun\right) + 1.61,
\end{equation}
where $\Lsun=3.84\times 10^{26}$~erg\,s$^{-1}$.
\revise{The upward spread due to high optical depths} barely affects the fit \revise{because optically thin cases are much more frequent for a uniform sampling of \MP and \Mdot}.
The formal errorbars are $\sigma_a=0.006$ and $\sigma_b=0.04$,
with the root-mean square residual $\mathrm{rms}=0.11$~dex,
but the half-spread at a given \LHa is %
$\sigma\approx0.3$~dex (shaded region in Figure~\ref{fig:Lacc_LHa}).
We recommend using $\sigma=0.30$ as the uncertainty when determining \Lacc from \LHa and propagating errors (Appendix~\ref{sec:sigma}).

We compare our fit to two empirical relationships derived for stars.
The blue region in Figure~\ref{fig:Lacc_LHa} shows the relationship of \citet[][hereafter \citetalias{Rigliaco+2012}]{Rigliaco+2012}.
To explain a given \LHa, \citetalias{Rigliaco+2012}'s fit requires an \Lacc smaller than our estimate by up to two orders of magnitude.
\revise{Since, in the shock model, the accretion energy is partitioned into the \Lya emission more by a factor of several tens than into the \Ha emission, 
such a high $\LHa/\Lacc$ ratio cannot be achieved. 
In \revise{contrast, in}
the magnetospheric accretion model
\revise{used for interpreting the}
\Ha emission in the stellar regime, \Lya emission should be less efficient than in the planetary shock model due to, for example, large optical thickness.}
\revise{Also, we note that a part of stellar \Ha\ energy (i.e., the energy heating the accretion column) might come from the stellar interior energy through the magnetic field, in addition to the accretion energy.}
\revise{\citetalias{Rigliaco+2012}'s fit}
also differs greatly in slope from ours. 
This is because \citetalias{Rigliaco+2012}'s empirical relationship was derived using stellar objects of higher $\Mdot$ (or \Lacc) than studied here;
for more massive stars, more energy is emitted in the UV continuum rather than in optical lines such as \Ha \citep{Zhou2014}. 
This is also the reason why the \Lacc estimated in \citet{Zhou+2021} via UV continuum is lower than ours. For stellar cases with stronger shock, \Lya should be much weaker than UV continuum and negligible, and \Lacc is well estimated only via UV continuum. But in the planetary shock emission, \Lya carries a large fraction of the energy, while this line is not observable due to strong circumstellar extinction.
\srevise{Even at TW Hya, one of the closest young stellar objects, the interstellar hydrogen column density is $\approx10^{19.5}$~cm$^{-2}$ \citep{Herczeg+2004}. Combined with the narrowness of the planetary \Lya line for low planet masses (full width at half maximum $\sim0.3$~\AA), this lets at most a percent of the \Lya reach us \citep{Landsman+Simon1993}. Thus, the flux ratio of \Lya to \Ha is around 0.1 or likely less, even at such a favourable target. For an object of $\gtrsim 20$ MJ, roughly ten times more \Lya passes through the ISM due to 
the greater line width.}

Also, for $\LHa\lesssim10^{-6}~\Lsun$,
\citetalias{Rigliaco+2012}'s extrapolated fit suggests $\LHa > \Lacc$: more energy is emitted in \Ha than is brought in by the accreting gas. This is not necessarily unphysical, since the \Ha does not have to originate from the accretion shock, but seems unlikely since here the only other energy source is the interior luminosity (usually smaller than \Lacc; \citealp{Mordasini+2017}).  %

The gray band in Figure~\ref{fig:Lacc_LHa} is from \citet[][hereafter \citetalias{Alcala+2017}]{Alcala+2017}, who extended the sample of \citet{Alcala+2014}. 
Since \citetalias{Alcala+2017} fit only to very-low-mass stars, their slope
should apply to planets presumably better more 
than \citetalias{Rigliaco+2012}. However, also \citetalias{Alcala+2017}'s fitted line differs from ours
by an order of magnitude. This reflects the
contrasting  %
\Ha emission mechanisms. Our model calculates \Ha from the shock-heated gas, while the stellar \Ha is thought to come (mainly) from an accretion funnel \citep{Hartmann+2016}.
Section~\ref{sec:stellar} discusses this more extensively.

%
Our \Lacc--\LHa relationship \revise{yields a higher \Lacc, for an observed \LHa, than both}
\citetalias{Alcala+2017}, $y=1.13x+1.74$, and shallower \citetalias{Rigliaco+2012}, $y=1.49x + 2.99$ \revise{do}.
Our curve also lies above those stellar correlations.
Therefore, a measurement of (or upper limit on) \LHa corresponds to a much higher accretion luminosity than inferred from the stellar fits.
Since \Mdot is unknown within several orders of magnitude, whereas the mass and radius uncertainties are much smaller, \Lacc should be set mostly by \Mdot. 
Thus, an observed $\LHa$ corresponds to a higher $\Mdot$ than inferred previously, suggesting that only strong accretors produce \Ha bright enough for detection.
This might help explain the low yields of recent \Ha surveys \citep{Cugno+2019,Zurlo+2020,Xie+2020}.

In Figure~\ref{fig:Lacc_LHa}, a range of \LHa values is covered both by the \citetalias{Alcala+2017} data and our model points (especially for $\ffill\gtrsim0.1$), at $\LHa\gtrsim5\times10^{-7}~\Lsun$.
The two %
are separated by %
1--2.5~dex at a given \LHa. However, the emission mechanisms %
likely differ  %
(Section~\ref{sec:modelcomp}). Therefore, the two relationships need not match. Also, if at a given \Lacc there are contributions from the shock and the accretion column,
the latter probably dominates at high \Mdot (Section~\ref{sec:compMdot}).
If however the temperature in the accretion column is below $T\approx10^4$~K or \Mdot is low, the surface shock more likely dominates.

Figure~\ref{fig:Lacc_LHa} shows the observed \LHa value of \PDSb
(vertical dotted line;
\srevise{\citealp{Zhou+2021}})
as a typical planetary \LHa (cf.\ \PDSc and Delorme~1 (AB)b; \citealp{Haffert+2019,Eriksson+2020}).
Our fit implies 
\srevise{$\Lacc\approx5\times10^{-5}~\Lsun$,}
which is respectively \srevise{about} ten and 100~times larger than for \citetalias{Alcala+2017} and \citetalias{Rigliaco+2012}, with the latter in the $\LHa>\Lacc$ region\footnote{%
When \citet{Wagner+2018} used \citetalias{Rigliaco+2012}'s fit, \Lacc was larger than \LHa because they estimated $\LHa=1.4\times10^{-6}~\Lsun$ \citepalias{Aoyama21}.
}.
Our predicted \Lacc is a lower limit if, as \citet{Hashimoto+2020} infer, there is extinction.

\subsection[Comparison of Mdot--LHa relationships]{Comparison of \Mdot--\LHa relationships}
  \label{sec:compMdot}

\begin{figure}
\begin{center}
    \includegraphics[width=0.47\textwidth]{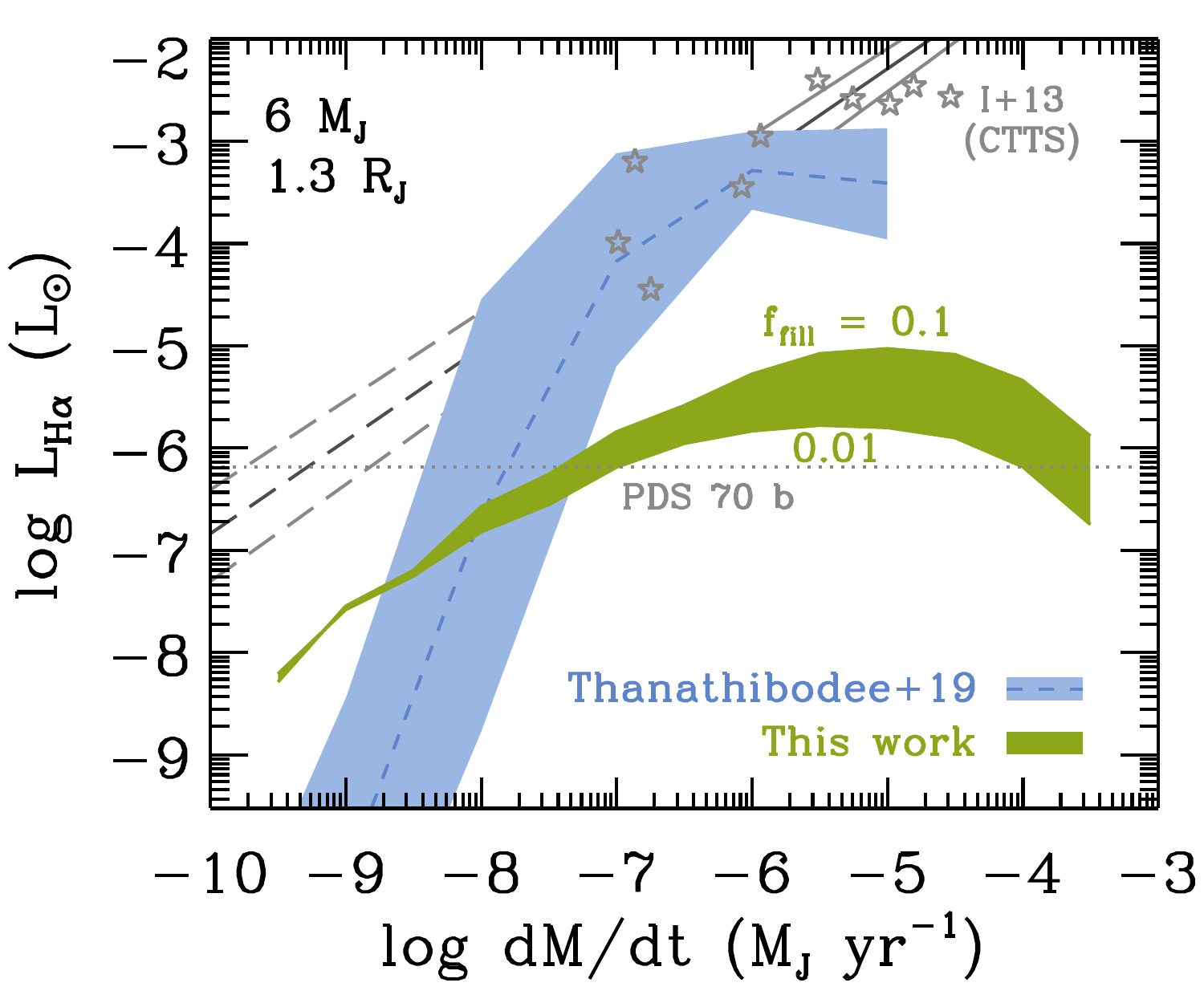}
    \caption{Restricted comparison of the
    $\Mdot$--$\LHa$
    relations
    for fixed $\MP=6~\MJ$
    and $\RP=1.3~\RJ$ (only here)
    from
    \citet[][blue band: 1~$\sigma$]{Thanathibodee+2019}
    and this work (green band).
    The bottom (top) border of our region is for $\ff=0.01$ (0.1), with \citetalias{Thanathibodee+2019} finding $\ff=0.01$--0.1.
    Gray stars and lines show the \citet{Ingleby+2013} data and fit ($\pm\sigma$) for CTTSs ($M\approx\Msun$) for comparison.
    \PDSb is highlighted (gray dotted line;
    \srevise{\citealp{Zhou+2021}}).
    For a version of this figure with extinction,
    see \citet{maea21}.
    }
    \label{fig:Mdot_LHa}
\end{center}
\end{figure}

A common approach in the literature is to use empirical \Mdot--\LHa correlations to infer \Mdot. This approach hides the possibly complex dependence of \LHa on several parameters into a best-fitting coefficient. Nevertheless, it is useful because \Mdot can vary by several orders of magnitude between objects while $\MP/\RP$ by much less, with $\LHa$ correlating with  %
$\Lacc\propto \MP\Mdot/\RP$ roughly.

We present in Figure~\ref{fig:Mdot_LHa} our $\Mdot$--$\LHa$ relationship.
To compare to \citetalias{Thanathibodee+2019},
we fix $\MP=6~\MJ$ and $\RP=1.3~\RJ$.
Since \citetalias{Thanathibodee+2019} assume magnetospheric accretion, we take $\ff=0.1$ and~0.01.
Figure~\ref{fig:Mdot_LHa} shows that the dependence of \LHa on \Mdot differs greatly between the surface shock (this work) and the heated accretion column  %
\citepalias{Thanathibodee+2019}. The latter has some overlap with the Classical T~Tauri Stars (CTTSs) relationship of \citet{Ingleby+2013}, who fitted higher-mass objects than \citetalias{Rigliaco+2012}. Both in \citetalias{Thanathibodee+2019} and here no extinction is considered. 
\revise{In our shock model, the emitted \LHa\ less steeply depends on the \Mdot\ relative to the model of \citetalias{Thanathibodee+2019}.}

The key point of Figure~\ref{fig:Mdot_LHa} is that the emission from the heated accretion column and that from the surface shock dominate in different regimes. Below a crossover value $\Mdotcr$,  %
the surface shock yields most of the \Ha luminosity, whereas at $\Mdot\gtrsim\Mdotcr$ the accretion column dominates.
The \citetalias{Thanathibodee+2019} models were fitted to a specific observation of a single object,
and we too considered only one $(\MP,\RP)$ combination here. While \Mdotcr presumably depends on these parameters,
the qualitative result that there is an \Mdotcr should be general.

If planets accrete mostly at $\Mdot >\Mdotcr$, the surface shock will not dominate \LHa. 
However, it is unclear how high $\Mdot/\Mdotcr$ can be.
For \PDSb, \citetalias{Thanathibodee+2019} %
fit a maximum temperature in the accretion column $\Tmax\approx10^4$~K.
If $\Tmax$ is significantly lower
\srevise{because accretion is less energetic for planets\footnote{\srevise{Even in energetic accretion by
massive objects ($\gtrsim 20M_\mathrm{J}$),
the shock still generates hydrogen lines. However, it hardly influences \LHa because most energy is emitted at shorter wavelengths (UV and X-ray).}},}
\Mdotcr could become very high and thus irrelevant in practice \revise{and the shock emission \srevise{would} dominate the \LHa}. Also, it has not (yet) been shown that magnetospheric accretion onto planets can occur at all. 
Finally, if planets have short phases of high \Mdot \citep[e.g.,][]{lubow12,Brittain+2020} but accrete mostly at low $\Mdot<\Mdotcr$ 
\citep[e.g.,][]{Tanigawa+2007}, 
observing them in a phase when the surface shock dominates is \revise{more likely}.

That both models cross
\srevise{near}
the \LHa of \PDSb seems fortuitous. 
The only other securely detected \revise{accreting} planetary-mass objects, \PDSc and Delorme~1~(AB)b, are fainter \citep{Haffert+2019,Eriksson+2020}.
Thus no planetary-mass observation has yet probed the regime where emission from an accretion column would clearly dominate.

\subsection[Which model is appropriate to estimate Mdot from LHa?]{Which model is appropriate to estimate \Mdot\ from \LHa?}
For planetary-mass objects, neither the shock model nor the magnetospheric accretion model indicate that the empirical relationships derived for accreting stars are applicable for planetary-mass objects. Thus, \Mdot\ should rather be estimated by using the relationship presented here or the modeling of planetary magnetospheric accretion as \citetalias{Thanathibodee+2019} did.
As discussed in Section~\ref{sec:compMdot}, for lower \Mdot (or lower \LHa), the shock-induced emission dominates over that from the magnetospheric accretion, and \PDSb is located 
\srevise{near} the threshold.
As of writing, surveys have found no planetary \Ha\ other than \PDSb \srevise{and c} \citep{Zurlo+2020, Xie+2020}. 
\srevise{When \Ha emission that was not detected due to its faintness is finally detected,}
we recommend using our relationship for such lower \LHa.
For \PDSb\ or planets as bright in \Ha\ as \PDSb, we discuss the way to distinguish the \Ha\ source in Section~\ref{sec:stellar}.

\section{Discussion: comparison of emission models}
 \label{sec:modelcomp}

In this section, we discuss which source of \Ha\ is expected for different assumptions, taking the planetary-surface shock %
as the fiducial case, and address how these mechanisms can be distinguished.
Appendix~\ref{sec:Zhu} reviews the upper limit on \LHa by \citet{Zhu2015}.

\subsection[H alpha from accretion funnel]{\Ha from accretion funnels}
  \label{sec:stellar}
  
An accretion shock is a general and efficient way to heat gas.
However, stellar-mass objects
have a large free-fall velocity $\vf\approx300~\kms$, which leads to too strong a shock for hydrogen-line emission \citep{Hartmann+2016}. Indeed, the
shock-heated gas reaches $T\gg10^5$~K,
stifling
significant line emission:
neutral hydrogen is rare and frequent electron--neutral collisions prevent (hydrogen-line-emitting) radiative cascades.
Also, the observed stellar \Ha %
line
is wide and comparable with the free-fall velocity. %
Since an unrealistically high temperature ($T\sim10^7$~K) would be required to explain this width by thermal broadening, the \Ha is thought to come from the accreting gas.
Namely, strongly magnetized protostars create an inner cavity in their protoplanetary disk and funnel the accreting gas along magnetic field lines \citep[e.g.,][]{garc20}, with a velocity distribution from $-\vf$ to $+\vf$ (back to front side; \citealp{Koenigl1991}).  %
Combined with appropriate radiative structures and inclinations, the mechanical Doppler shift from this velocity distribution
can reproduce observed \Ha\ widths
\citep{Hartmann+1994}.

On the other hand, $\vf$ for planets is much 
\revise{lower} than for stars, with $\vf\approx100\,\sqrt{M_5/R_2}~\kms$ (where $X_n\equiv X/[n~X_{\mathrm{J}}]$) for planets. Such a shock generates a propitious  %
environment ($T\sim10^5$~K) for \Ha emission \citepalias{Aoyama+2018} and can reproduce the observed \Ha\ line width \citepalias{Aoyama+Ikoma2019}. Therefore, for planetary \Ha, the shock-heated gas is a strong candidate, unlike in the stellar case.

\citetalias{Thanathibodee+2019} modeled the \Ha from a planetary accretion funnel by extending models of stellar  %
emission \citep{Muzerolle+2001}. If such accretion funnels exist for planets, the free-falling gas will be shock-heated at the planetary surface and emit non-negligible \Ha.
Then, the observed \Ha should be a mixture of two components: from the funnel and from the shock.

Non-Gaussian line wings would suggest
a contribution from funnels.
In funnel emission, the line broadening
comes from
the bulk (not thermal) velocity. Therefore, the line is a superposition of narrow ($T\approx10^4$\,K) Gaussians and therefore not necessarily Gaussian.
For shock emission, the post-shock gas exhibits a range of temperatures and velocities. Thus, the profile is also a superposition of Gaussians, but each component is much wider
($T\gtrsim 10^5$\,K). Also, the velocity change in the emitting layers is much less than the highest thermal velocity that determines the widest profile.
Therefore, the line is nearly Gaussian,
especially in the wings. Self-absorption likely makes the \Ha line center non-Gaussian \citepalias{Aoyama+2018}, but optically thinner lines from shocks could be 
completely Gaussian. From funnels, also the thinner lines are non-Gaussian.

The asymmetry across the line center can help to distinguish shock from funnel emission.
The shock emission necessarily has a wider red-side profile because of the receding emitting gas, while the funnel emission is freer and can have a broader blue side.
Distinguishing this %
requires resolving the line
($R\gtrsim10^4$).

Also, it is difficult to make the accretion funnel hot enough to produce observable \Ha\ even in stellar cases \citep{Martin1996}, and the heating mechanism, possibly magnetic in nature, remains an open question. Accordingly, \citetalias{Thanathibodee+2019} used a parametrized temperature structure.
For young, luminous planets, even though the \citet{Christensen+2009} scaling predicts a strong magnetic field, the accretion funnel could have a lower temperature (due somehow to the shallower potential) and emit weaker \Ha\ than in the stellar case.
Thus the emission of \Ha by planetary accretion funnels is an interesting but currently difficult-to-assess possibility.

\subsection[H alpha from a strong shock on CPD surface]{\Ha from a strong shock on the CPD surface}
  \label{sec:Sshock}
  
The gas that enters the inner parts of the planetary Hill sphere falls onto the circumplanetary disk (CPD) roughly vertically (see e.g.\ \citealp{Schulik+2019,Schulik+2020} for $\MP\lesssim1~\MJ$).
Three-dimensional isothermal hydrodynamic simulations showed that the velocity just above the CPD surface is comparable to the free-fall velocity  %
\citep[e.g.,][]{Tanigawa+2012}. Therefore, the shock-heated gas can get hot enough to emit observable \Ha (\citealp{Szulagyi+Mordasini2017}; \citetalias{Aoyama+2018}).
However, in this case, most of the gas falls far from the planet, where %
\Ha emission hardly occurs due to the low \revise{free-fall} velocity \revise{of the accreting gas} (\citetalias{Aoyama+2018}; Section~5.2 of \citetalias{Aoyama21}). Thus, only a small fraction of the accreting gas can contribute to the \Ha\ emission.

Consequently, if %
the gas entering the Hill sphere undergoes
shocks on both the CPD and the planet, %
the former
is likely negligible, unless only a small fraction hits the planet.
If the CPD connects continuously to the planetary surface \citep{Owen+Menou2016,Dong+2020} and/or the CPD gas flows outward rather than towards the planet \citep{Szulagyi+2014}, \revise{existing models show that} the CPD surface shock would be the only \Ha\ source because there would be no strong shock on the planet.  %
However, to reproduce a given \Ha luminosity, the CPD surface shock requires a higher mass influx rate\footnote{As \citetalias{Szulagyi+Ercolano2020} write, the planet might not accrete all the mass inflowing towards the CPD, leading to the distinction between ``influx'' and ``accretion''.} onto the CPD than the planetary surface shock case\revise{, because most of accreting gas hardly contribute to the \Ha flux}.
If the shock is at a large distance above the planet, 
no strong shock is predicted (see \S~\ref{sec:wshock}).

In the CPD shock case, the \Ha\ spectral profile is similar to the case of the planetary surface shock,
and it is hard to distinguish the two cases with current instrumental resolution (e.g., $R\approx2500$ with MUSE/VLT).
Instead, most of the gas undergoes a weak shock in the far regions of the CPD, and there is much more cool ($<10^4$~K) gas, which emits molecular lines, than in the case of the planetary surface shock.
Also, the higher mass influx %
needed to reproduce a given \LHa should lead to a higher temperature for the planetary photosphere and the CPD midplane, making both easier to observe.

\subsection[H alpha from an extremely large planet]{\Ha from an extremely large planet}
  \label{sec:wshock}
Some global 3D radiative-hydrodynamic simulations for $\MP\sim1$--$10~\MJ$ have obtained a roughly spherically symmetric accretion front 
$\approx 55~\RJ$
large in radius
\citep{Szulagyi+Mordasini2017}.  
Interpreted as the planet radius, this size is unexpected at those masses of several Jupiter masses in classical planet modeling.
With the density--temperature structure around the gravitational-potential point mass from their simulation,
\citetalias{Szulagyi+Ercolano2020} integrated the radiative-transfer equation, using the \citet[hereafter \citetalias{storey95}]{storey95} emissivities in the source term and the gas and dust opacity in the absorption term. This yielded hydrogen-line luminosities.

We discuss 
two \revise{critical} aspects of %
\citetalias{Szulagyi+Ercolano2020}'s approach:
\begin{itemize}

\item[(i)] The use of \citetalias{storey95}. This model was originally derived in the context of photoionization by, e.g., Wolf--Rayet stars. As detailed in \citetalias{Aoyama21},
these tables do not apply here mainly because they 
neglect the ground-state population and, therefore, collisional excitations from \srevise{that} state. Especially for a moderate shock (e.g., Figures~2 and~3 in \citetalias{Aoyama+2018} shows the case of $v_0=40~\kms$), the low ionization fraction makes the ground state be the most populated state. This contradicts the assumption of \citetalias{storey95} (see also Section~4 in \citealp{Hummer+Storey1987}
).
Thus, %
line emissivities based on \citetalias{storey95}
differ fundamentally
from the ones %
from a direct non-equilibrium calculation.

\item[(ii)] The thickness of the cooling region. For relevant densities,
its thickness 
in our models
is (much) less %
than the
planetary radius ($\sim10^{10}$~cm), as  
expected.
For example, in Figure~6 of \citetalias{Aoyama21}, the characteristic thickness is $10^7$~cm. Currently, our model does not include all coolants, in particular metals;
if we did, %
the region would become even thinner \citep{Aoyama+2018,Aoyama21}. In \citetalias{Szulagyi+Ercolano2020}, however, the grid cells at the shock are at least of order $\sim\RJ$,
much larger than the physical size of the cooling region. Since the %
emission of a cell is the product
of its volume and emissivity, which depends strongly on temperature,
the size of the highest-$T$ cells directly  influences \citetalias{Szulagyi+Ercolano2020}'s predicted line intensities. This might explain why, in
some of their %
cases,  %
the \Ha line luminosity is much larger than the total accretion luminosity ($\LHa\gg\Lacc$). \revise{This means the radiative cooling and emission are not consistently treated.}
\end{itemize}

These points demonstrate that the hydrogen-line emission from a planetary shock cannot be calculated by combining \citetalias{storey95} with the output of %
radiation-hydrodynamical simulations.
\revise{Especially concerning the thickness of the cooling region, this approach could keep the gas temperature high for longer than in reality and lead to overestimate of the line luminosity.}
This holds for \citetalias{Szulagyi+Ercolano2020} even though their simulations are highly resolved for a global 3D simulation with an impressive dynamic range 
of $\sim10^4$ in lengthscale and thus capture the general dynamics. The issue is that 3D simulations necessarily remain low-resolution compared to the postshock cooling, which acts on a very different (microphysical) scale than the hydrodynamical processes. This challenge holds for the 1D simulations of \citet{Marleau+2017,Marleau+2019b} too, despite their higher resolution.

A shock at tens of $\RJ$ is %
distinguishable spectroscopically, assuming that any \Ha is emitted (which requires $\vf\gtrsim30~\kms$; \citetalias{Aoyama+2018}).
The %
\Ha profile is  %
narrower than in the other cases (Sections~\ref{sec:stellar} and~\ref{sec:Sshock})
because the gas is slower than %
in an accretion funnel in which the gas accelerates until the planet's surface at a few $\RJ$,
and cooler than when heated by a strong shock, for which $T>10^5$~K.
The half-width at half-maximum
is narrower than the shock velocity $\vf\leqslant35\,\sqrt{M_{10}/R_{30}}~\kms$ because the infall is supersonic. Also, the photospheric component has a lower effective temperature $\Teff\propto\RP^{-1/2}$.

Could this apply to \PDSb? \citetalias{Haffert+2019} reported a spectral width slightly above $100~\kms$. 
The \Ha from a weak shock is at least three times narrower, which seems inconsistent with these observations, but the measured width might be overestimated because it is comparable to the instrumental resolution (\citetalias{Thanathibodee+2019}; \citealp{Hashimoto+2020}).
However, \citet{wang21vlti} obtained a photospheric radius $\RP\approx2~\RJ$. Therefore, a very weak shock seems unlikely for \PDSb.

\section{Summary and discussion}
 \label{sec:disc}
We have considered the predictions of the \Ha flux from sophisticated non-LTE models of the postshock emission from \citet{Aoyama+2018} as applied to the scenario that the shock occurs on the planet surface, as in \citet{Aoyama+Ikoma2019}. Using a broad range of \Mdot and \MP relevant for forming planets, we have shown for the first time the \Lacc--\LHa relation for the planetary-surface shock, comparing to previously-used stellar relationships (Figure~\ref{fig:Lacc_LHa}). Appendix~\ref{sec:morelines} extends this to other hydrogen lines. We then
compared our \Mdot--\LHa relationship to
that of \citet{Thanathibodee+2019}.
Finally, we %
put in perspective
accretion contexts that can lead to \Ha emission (Section~\ref{sec:modelcomp}).

In summary:
\begin{enumerate}
\item The relationship
$\log (\Lacc/\Lsun) = 0.95 \log (\LHa/\Lsun)+1.61$
(Equation~(\ref{eq:Lacc_LHa}))
is markedly higher at a given \LHa  %
than extrapolating the stellar relationships
from \citet{Rigliaco+2012} and \citet{Alcala+2014,Alcala+2017}.
Thus \Ha production is less efficient for planets (Figure~\ref{fig:Lacc_LHa}).
\item For magnetospheric accretion, the contribution of heated accretion columns \citep{Thanathibodee+2019} dominates at high \Mdot, 
whereas %
the surface-shock contribution is larger at low \Mdot.
Whether for realistic \Mdot the emission from the column will ever dominate, however, depends on the highly uncertain temperature in that model,
and presumably %
also on mass and radius.  %
\PDSb happens to be in the intermediate-\Mdot regime (Figure~\ref{fig:Mdot_LHa}), if the accretion funnels %
are
hot enough to emit \Ha.
\item 
A non-Gaussian \Ha wing or a wider profile on the blue side
indicates that a hot accretion funnel \citep{Thanathibodee+2019} contributes to the line, in addition to the shock-heated gas on the planetary surface \citep{Aoyama+Ikoma2019,Aoyama21} (Section~\ref{sec:stellar}).
A weak shock on a large planet (tens of $\RJ$)
should have a narrow line (\S~\ref{sec:wshock}).

\item Importantly, we have argued (\S~\ref{sec:wshock}) that the hydrogen-line emission from  %
large planets cannot be calculated by applying \citet{storey95} on the output of LTE, relatively low-resolution (compared to the disequilibrium microphysical processes in the postshock cooling region) radiation-hydrodynamical simulations such as those of \citet{Marleau+2017,Marleau+2019b} or \citet{Szulagyi+Ercolano2020}.
\end{enumerate}

The new \Lacc--\LHa relationship has important implications. One
is that \PDSb is now predicted to have
\srevise{$\Lacc\approx5\times10^{-5}~\Lsun$}
(see Figure~\ref{fig:Lacc_LHa}) instead of ten to 100 times smaller using the extrapolated stellar relationships.
Also, \citet{Zurlo+2020} reached an average \Ha upper limit of $\LHa \approx 5\times10^{-7}~\Lsun$ beyond $\approx0\farcs1$ %
in their survey. Using the \citet{Rigliaco+2012} relationship, this would translate to $\Lacc<4\times10^{-7}~\Lsun$, while we find instead $\Lacc<3\times10^{-5}~\Lsun$, a much
looser constraint.
Finally, \citet{Close2020} estimated the future observability of \Ha-emitting planets but based on the \citetalias{Rigliaco+2012} scaling. Using instead ours, we estimate from his Figure~8 that a large fraction of the planets should remain detectable thanks to the high assumed $\Mdot$, where both scalings differ only by $\lesssim1$~dex.

Finally, some words about extinction.
Apart from the ISM, the matter either in the accretion flow onto the planet or in the PPD layers above the planet
can contribute to the extinction.
\citet{Szulagyi+2019II} and \citet{Sanchis+2020} argued that extinction by circumstellar and circumplanetary materials could make planets or their CPDs more challenging to detect. This seems qualitatively realistic, but the extent depends strongly on the details of the accretion flow,
which are heavily influenced by the numerical resolution, and on the uncertain dust properties.

We did not consider extinction by the gas nor the dust 
around the planet.
This should be justified towards low \Mdot, and for the dust it will hold especially if accreting planets are found in gaps \citep{Close2020},
where  %
the local dust-to-gas ratio is much lower than the global average
%
\citep[e.g.,][]{dr19}.
%
%
%
\revise{Since extinction decreases the observed flux, the true \LHa is higher than the \LHa estimated from the observed flux. Therefore,}
our relationship is robustly a \textit{lower bound} on the \Lacc implied by \revise{the observed \Ha flux.}
Depending on the details of the accretion and viewing geometries,
heavy extinction could be avoided. To assess this observationally, comparing theoretical predictions of line ratios to simultaneous observations of several accretion tracers \citep{Hashimoto+2020} seems a promising avenue.

\section*{Acknowledgements}

We thank C.~Manara and S.~Edwards for very informative discussions.
Parts of this work were conducted during the visit of YA as a Visiting Scholar of the SPP~1992 program of the Deutsche Forschungsgemeinschaft (German Science Foundation; DFG) and also of the JSPS Core-to-Core Program ``International Network of Planetary Sciences (Planet$^2$).'' YA and MI acknowledge the support from JSPS KAKENHI grants Nr.~17H01153 and 18H05439.
G-DM acknowledges the support of the DFG priority program SPP~1992 ``Exploring the Diversity of Extrasolar Planets'' (KU 2849/7-1 \revise{and MA~9185/1-1}) and support from the Swiss National Science Foundation under grant BSSGI0$\_$155816 ``PlanetsInTime''.

\appendix

\section{Errorbars on the relationships}
 \label{sec:sigma}

The formal statistical errorbars on the fit parameters $a$ and $b$ are usually taken to derive errorbars on the derived \Lacc (or \Mdot; see below). For a general fit $\log\left(\Lacc/\Lsun\right) = a\log\left(\LHa/\Lsun\right) + b$, the spread $\siglgLacc$ for the underlying distribution of parameters is given by the standard propagation of errors:
\begin{equation}
\label{eq:sigLL}
 \siglgLacc = \sqrt{ \sigma_a^2\log(\LHa)^2 + \sigma_b^2 + a^2 \siglgLHa^2 },
\end{equation}
where $\sigma_{a,~b,~\LHa}$ are respectively the uncertainties on $a$, $b$, and $\LHa$.
With this, $\Lacc\times10^{\pm\siglgLacc}$ gives the 1-$\sigma$ range of values at a given \LHa.

The use of $\sigma_{a,~b}$ implicitly assumes that the underlying relationship between \LHa and \Lacc has no intrinsic spread, with some unknown, nuisance parameter(s) leading to noise in the `observed' (from data or models) \Lacc. Our $\sigma_{a,~b}$ are much smaller than those of the literature relationships only because we use more model points for the fit than data points were used.
However, in reality the spread arises because both \LHa and \Lacc depend on a number of physical parameters (\Mdot, \MP, \RP) in general in a different way. Thus it would be more appropriate to use the spread of the points $\sigma$ than the formal error, contrary to what has been done up to now.

As an example, for the \citetalias{Alcala+2017} fit, the formal uncertainty on $\Lacc$ from $\log(\LHa/\Lsun)=-6.805\pm0.095$
(for \PDSb; \citetalias{Haffert+2019})
is $\sigform = 0.11$~dex, with the contributions from the formal errors on $a$ and $b$ dominating.
Meanwhile, the spread in the original data, which reaches down only to $\log(\LHa/\Lsun)\approx-6$, is rather $\sigma\approx0.5$~dex at the low end,
and mostly $\sigma\approx0.7$~dex over the whole range.  %
Thus using only the formal errorbars strongly underestimates the uncertainty in the derived $\Lacc$. The same conclusion is reached when considering \citetalias{Rigliaco+2012} and \citet{Alcala+2014}, for both of which the spread of $\Lacc$ is $\sigma\approx0.5$~dex over their range.

\section{A comment on Zhu (2015)}
 \label{sec:Zhu}
\citet{Zhu2015} presented an expression for the \Ha\ luminosity from accreting planets in the context of magnetospheric accretion (his Equation~(21)):
\begin{equation}
 \label{eq:zhu1522}
 \LHa = 4\pi \Rtrunc^2 \times \pi B_\nu (8000~\mathrm{K},\nuHa)\times\frac{\vf}{c} \nuHa,
\end{equation}
where $B_\nu(T,\nu)$ is the Planck function, $\Rtrunc$ is the magnetospheric truncation radius \citep{Koenigl1991},  %
$\nuHa$ is the \Ha frequency,
and $c$ is the speed of light.
This is meant not as a precise calculation but as a rough upper limit. Still, we comment on its applicability to put it in context.

Equation~(\ref{eq:zhu1522}) assumes that the surface of the magnetosphere is covered by an optically thick layer of hot ($T\approx8000$~K) gas, and that the atomic hydrogen level populations are in thermal collisional equilibrium, i.e., given by the Boltzmann distribution at the gas temperature $T$. This will likely hold since the free-fall time is long compared to the thermal timescale, so that the radiation and gas temperatures become equal. Then if the emitting region is optically thick, the sphere of radius $\Rtrunc$ emits \Ha following the Planck function at this $T$. The densities could be such that the gas is optically thick \citep{Zhu2015}, but a realistic geometry should lead to a smaller emitting area. Thus the first two factors of Equation~(\ref{eq:zhu1522}) are upper limits.

The last factor in Equation~(\ref{eq:zhu1522}) assumes that the line width is equal to the infall velocity,
as the magnetospheric accretion model \citep[e.g.,][]{Hartmann+1994} suggests, and assumes a top-hat line shape, i.e., that the gas is optically thick (so that the line is saturated) within $\sim\vf$ of $\nuHa$ and thin outside. 
While this line width is consistent with the velocity distribution of the accretion funnel, the width from each region is represented by the thermal Doppler width rather than $\vf$. Therefore, the Doppler width is more realistic, while $\vf$ is better to make sure the estimate is truly an upper limit.

In summary, Equation~(22) of \citet{Zhu2015} represents a very conservative upper limit to the \Ha flux expected from magnetospherically-accreting planets.
In our model, most \Ha emission does typically come from regions at $T\approx(1$--$2)\times10^4$~K. However, the \Ha is usually optically thin there, making Equation~(\ref{eq:zhu1522}) really an upper limit. Indeed, for the input grid of models in Figure~\ref{fig:Lacc_LHa}, it predicts $\LHa\lesssim10^{-3}~\Lsun$, independently of \Mdot. Comparing to Figure~\ref{fig:Lacc_LHa}, this certainly holds.

Finally, based on their non-detections, \citet{Zurlo+2020} derived an upper limit
on the planetary mass from the upper limit of $\vf$ derived with Equation~(\ref{eq:zhu1522})\footnote{The upper value of $T\sim10^8$~K quoted by \citet[][]{Zurlo+2020} above their
Equation~(2) for the shock temperature in \citetalias{Aoyama+2018} is only the \textit{non-equilibrium}
value in extreme cases in a thin layer} but not the gas temperature near the planet; for the latter, see Equation~(33) of \citet{Marleau+2019b}.
.
However, while Equation~(\ref{eq:zhu1522}) gives an upper limit of \LHa\ for a given planetary mass, the equation does not necessarily give an upper limit of planetary mass for a given \LHa.
It would be interesting to repeat their analysis using more detailed models.  %

\section{Correlation between the line and accretion luminosities for other lines}
 \label{sec:morelines}

As in Appendix~E of \citet{Alcala+2017}, we provide fits to the relationship between the line luminosity \Lline and the accretion luminosity \Lacc in our model for several hydrogen lines other than \Ha, including near-infrared lines. We consider \Lya, \Lyb, \Lyg, and the transitions up to an upper level $\nup=8$ in the Balmer (H), Paschen (Pa), and Brackett (Br) series. Given that we include in our model lines only up to $\nup=10$ \citep{Aoyama+2018}, these fluxes should be reliable. %
As in Equation~(\ref{eq:Lacc_LHa}), we write
\begin{equation}
\label{eq:Lacc_Lline}
 \log_{10}\left(\Lacc/\Lsun\right) = a \times \log_{10}\left(\Lline/\Lsun\right) + b.
\end{equation}
We use the same grid of $(\Mdot,\MP,\ffill)$ values as in Section~\ref{sec:compLacc},
and also perform straightforward least-squares fitting with \texttt{gnuplot}'s built-in \texttt{fit} function.
As for \Ha, we use for the fit for each line the points
at $\Lacc\leqslant10^{-4}~\Lsun$.
For the Lyman-series lines and the $\alpha$ lines of the other series,
this excludes the region with a large spread in \Lacc at a given \Lline.
For the other lines, which are optically thinner, this restriction does not change the fit much and only effectively adds some statistical weight to the lower luminosities.

\begin{deluxetable}{l r ccc m{0.1ex} ccl}
\tablecaption{Relationships between line and accretion luminosities}
\tablehead{%
\colhead{} & \colhead{} & \multicolumn{3}{c}{PMCs (this work)} & & \multicolumn{3}{c}{CTTSs}\\%
\cline{3-5} \cline{7-9}
\colhead{Line} & \colhead{$\lambda$}  %
& \colhead{$a$} & \colhead{$b$} & \colhead{$s$} &
& \colhead{$a$} & \colhead{$b$} & \colhead{$s$} \\[-0.2cm]
\colhead{} & \colhead{($\upmu$m)} & \colhead{} & \colhead{(dex)} & \colhead{(dex)} &  
& \colhead{} & \colhead{(dex)} & \colhead{(dex)}%
}
\startdata
  \Lya   &  0.121 &  0.90 & 0.43 & 0.24 &  & ---  & ---  & ---  \\
  \Lyb   &  0.103 &  0.86 & 0.83 & 0.21 &  & ---  & ---  & ---  \\
  \Lyg   &  0.097 &  0.86 & 1.17 & 0.21 &  & ---  & ---  & ---  \\
  \hline
  \Ha    &  0.656 &  0.95 & 1.61 & 0.11 &  & 1.13 & 1.74 & 0.41 \\
  \Hb    &  0.486 &  0.87 & 1.47 & 0.12 &  & 1.14 & 2.59 & 0.30 \\
  \Hg    &  0.434 &  0.85 & 1.60 & 0.14 &  & 1.11 & 2.69 & 0.29 \\
  \Hd    &  0.410 &  0.84 & 1.77 & 0.15 &  & 1.07 & 2.64 & 0.32 \\
  H\,7   &  0.397 &  0.83 & 1.91 & 0.15 &  & 1.06 & 2.69 & 0.32 \\
  H\,8   &  0.389 &  0.83 & 2.04 & 0.16 &  & 1.06 & 2.73 & 0.30 \\
  \hline
  \Paa   & 1.875  &  0.93 & 2.49 & 0.10 &  & ---  & ---  &  --- \\
  \Pab   & 1.282  &  0.86 & 2.21 & 0.12 &  & 1.06 & 2.76 & 0.45 \\
  \Pag   & 1.094  &  0.85 & 2.28 & 0.14 &  & 1.24 & 3.58 & 0.36 \\
  \Pad   & 1.005  &  0.84 & 2.38 & 0.15 &  & 1.22 & 3.74 & 0.40 \\
  Pa\,8  & 0.954  &  0.83 & 2.49 & 0.15 &  & 1.09 & 3.19 & 0.42 \\ %
 \hline
  \Bra   & 4.051  &  0.94 & 3.32 & 0.10 &  & 1.81 & 6.45 & 0.1  \\ %
  \Brb   & 2.625  &  0.87 & 2.88 & 0.12 &  & ---  & ---  & ---  \\
  \Brg   & 2.166  &  0.85 & 2.84 & 0.14 &  & 1.19 & 4.02 & 0.45 \\
  \Brd   & 1.944  &  0.84 & 2.88 & 0.15 &  & ---  & ---  & ---  \\
\enddata
\tablecomments{%
Coefficients pertain to $\log_{10}\left(\Lacc/\Lsun\right) = a \times \log_{10}\left(\Lline/\Lsun\right) + b$ as in Equation~(\ref{eq:Lacc_LHa}).
PMCs: planetary-mass companions.
CTTSs: Classical T Tauri Stars.
Air wavelengths are reported, except for Lyman lines (vacuum). %
The CTTS fits are from \protect\citet{Alcala+2017},
except for \Bra, from \protect\citetalias{Komarova+Fischer2020}.
The $s$ values are the standard deviations of the linear fits
(estimated by eye for \protect\citetalias{Komarova+Fischer2020}; $N=7$ data points). This is not the spread of the data, which is
for example $\pm\sigma=\pm0.3$~dex for our \Ha line and at most 0.5~dex for some of the other lines (see Figures~\ref{fig:morelines1}--\ref{fig:morelines3}).
}
\label{tab:allfits}
\end{deluxetable}

The fit coefficients are reported in Table~\ref{tab:allfits} and compared to the stellar case. Where available, the latter are from \citet{Alcala+2017} with the exception of \Bra, from \citet[][hereafter \citetalias{Komarova+Fischer2020}]{Komarova+Fischer2020}.
Our coefficients are mostly slightly sub-linear ($a\approx0.9$), with a flattening (smaller $a$) towards higher-energy transitions within each series. This holds also in the Balmer series for stars but not in the Paschen series.
The $\sigma$'s give the standard deviation of the model points (or the data, for the CTTS column) with respect to the fit, but note that the spread of the points is larger (see discussion in Appendix~\ref{sec:sigma}).

All these lines should trace accretion, contrary to the case for CTTSs, where other processes can alter several lines (including, in fact, \Ha, which led \citet{Alcala+2017} not to recommend it as an accretion tracer).  %
However, for any line to be observable as a shock excess, it must be stronger than the local (pseudo)continuum
if the observations do not resolve it,
or higher than the ``noise'' (i.e., the room-mean-square level) of the (pseudo)continuum for spectrally-resolved observations.
Being at short wavelengths $\lambda\approx300$--400~nm, lines such as \Hg and higher-order Balmer lines are difficult to observe with existing instruments %
but they are included for completeness.
The \Lya line is also not likely to be observed but is relevant in thermochemical models \citep[e.g.,][]{rab19}.  %
The James Webb Space Telescope (JWST) should observe \Bra as \citetalias{Komarova+Fischer2020} pointed out, and Integral Field Unit (IFU) of the planned
second-generation High Resolution Spectrograph (HIRES) on the Extremely Large Telescope (ELT), expected to come online in the next decade, will cover 1.0--1.8~$\upmu$m, which includes several of the other lines.
Its tremendous resolution of $R\approx100,000$--$150,000$ should allowed detailed studies of the kinematics of the infalling gas.

\begin{figure*}
\begin{center}
    \includegraphics[width=0.8\textwidth]{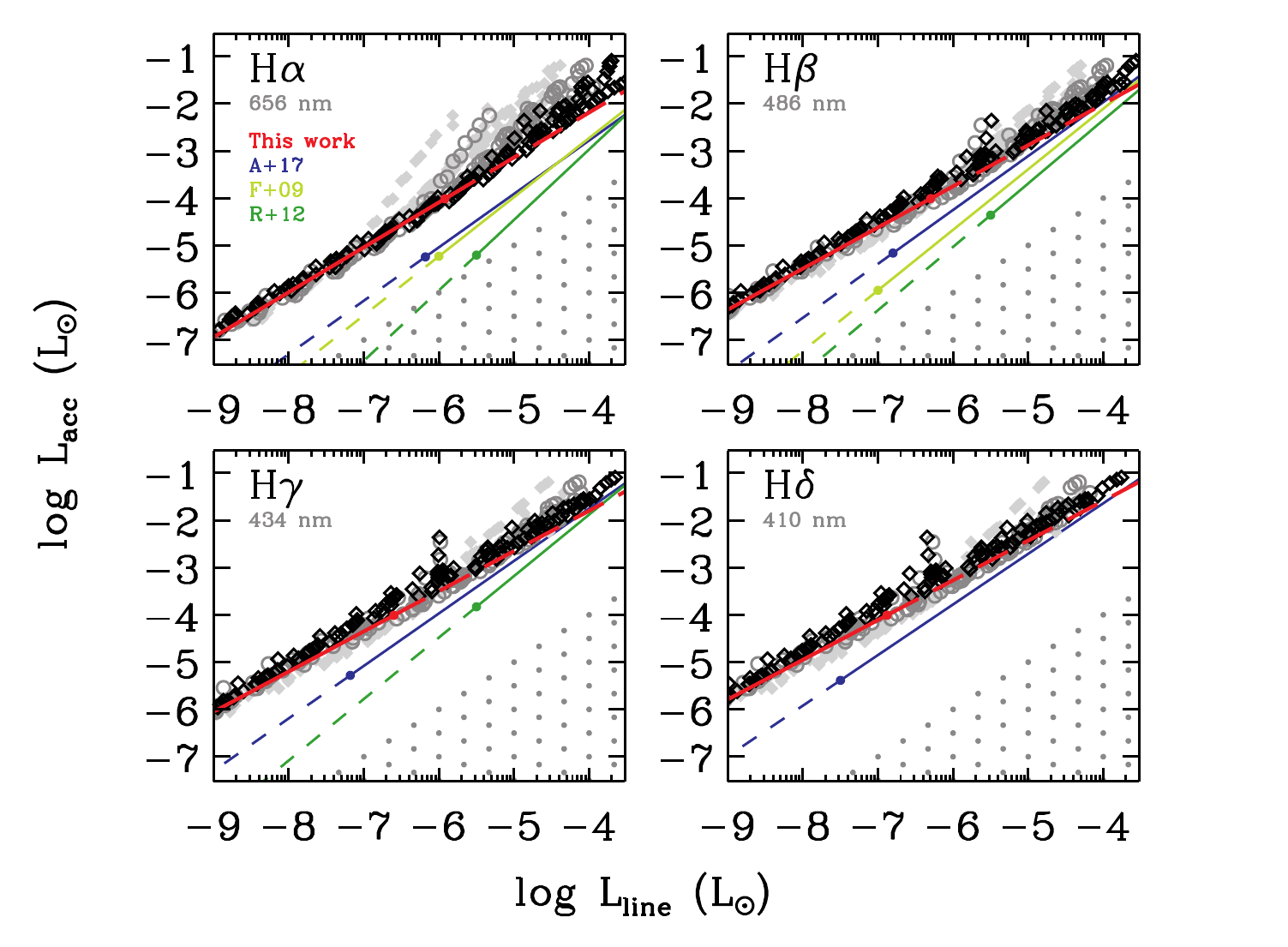}
    \includegraphics[width=0.8\textwidth]{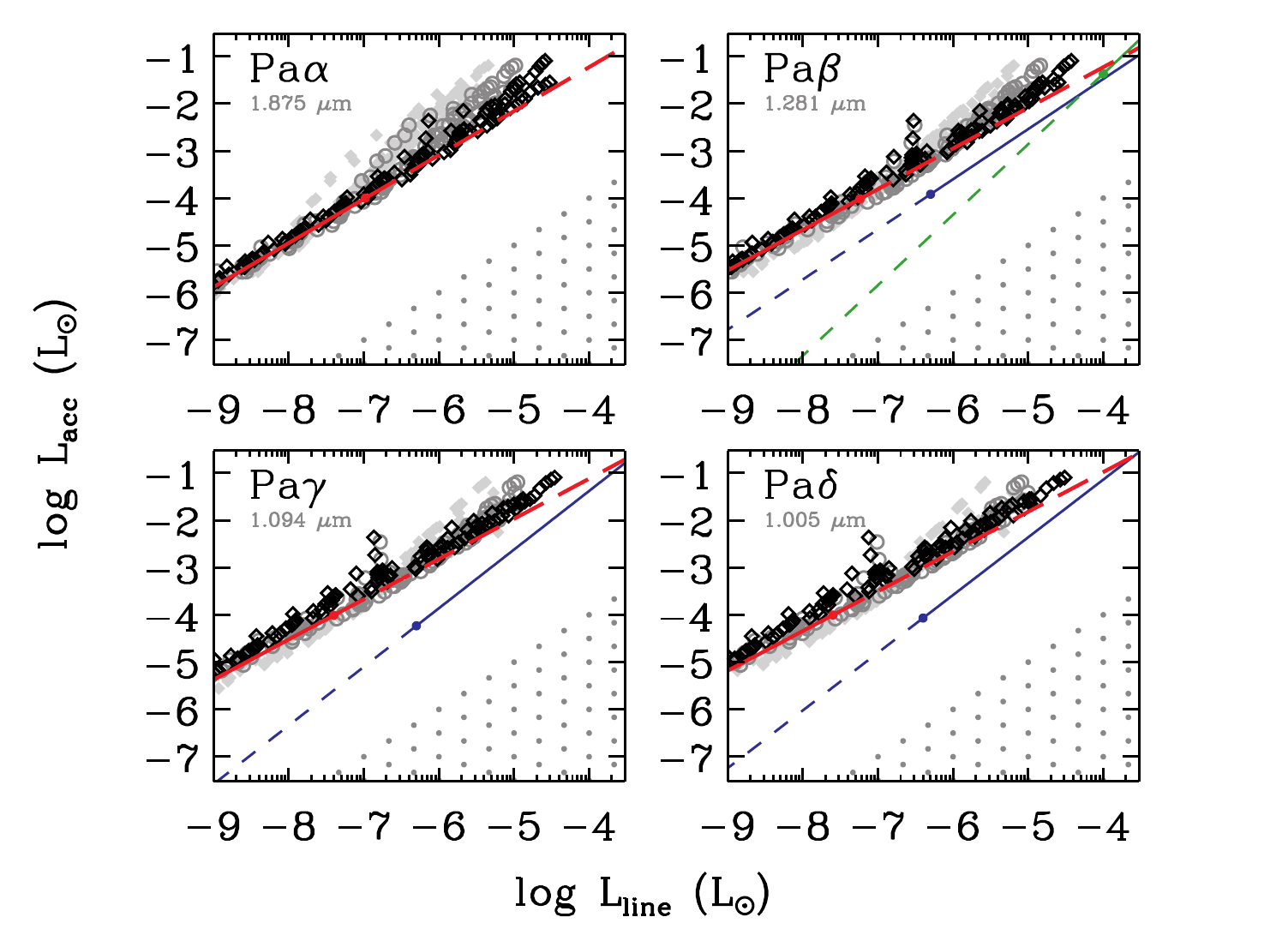}
    \caption{
Correlation between line and accretion luminosities for
hydrogen lines (see labels).
Symbols  %
show our model
for the same \Mdot and \MP as in Figure~\ref{fig:Lacc_LHa}
for $\ffill=1$, 0.1, and~0.01 (\textit{black, dark gray, and pale gray symbols} respectively) along with the fit (\textit{red line}; Table~\ref{tab:allfits}) from the points with $\Lacc\leqslant10^{-4}~\Lsun$ (solid).
Where available, the fits of
\citet[][\textit{blue}]{Alcala+2017},
\citet[][\textit{lime}]{Fang+2009}, and
\citet[][\textit{green}]{Rigliaco+2012}
are shown (solid: where their data exists; dashed: extrapolation).
A dot marks the transition between fits and extrapolations
but in our case the latter hold everywhere as a lower limit.
In the dotted region,  %
$\Lline>\Lacc$, which is likely unphysical (see Section~\ref{sec:compLacc}).
 }
    \label{fig:morelines1}
\end{center}
\end{figure*}

\begin{figure*}
\begin{center}
    \includegraphics[width=0.8\textwidth]{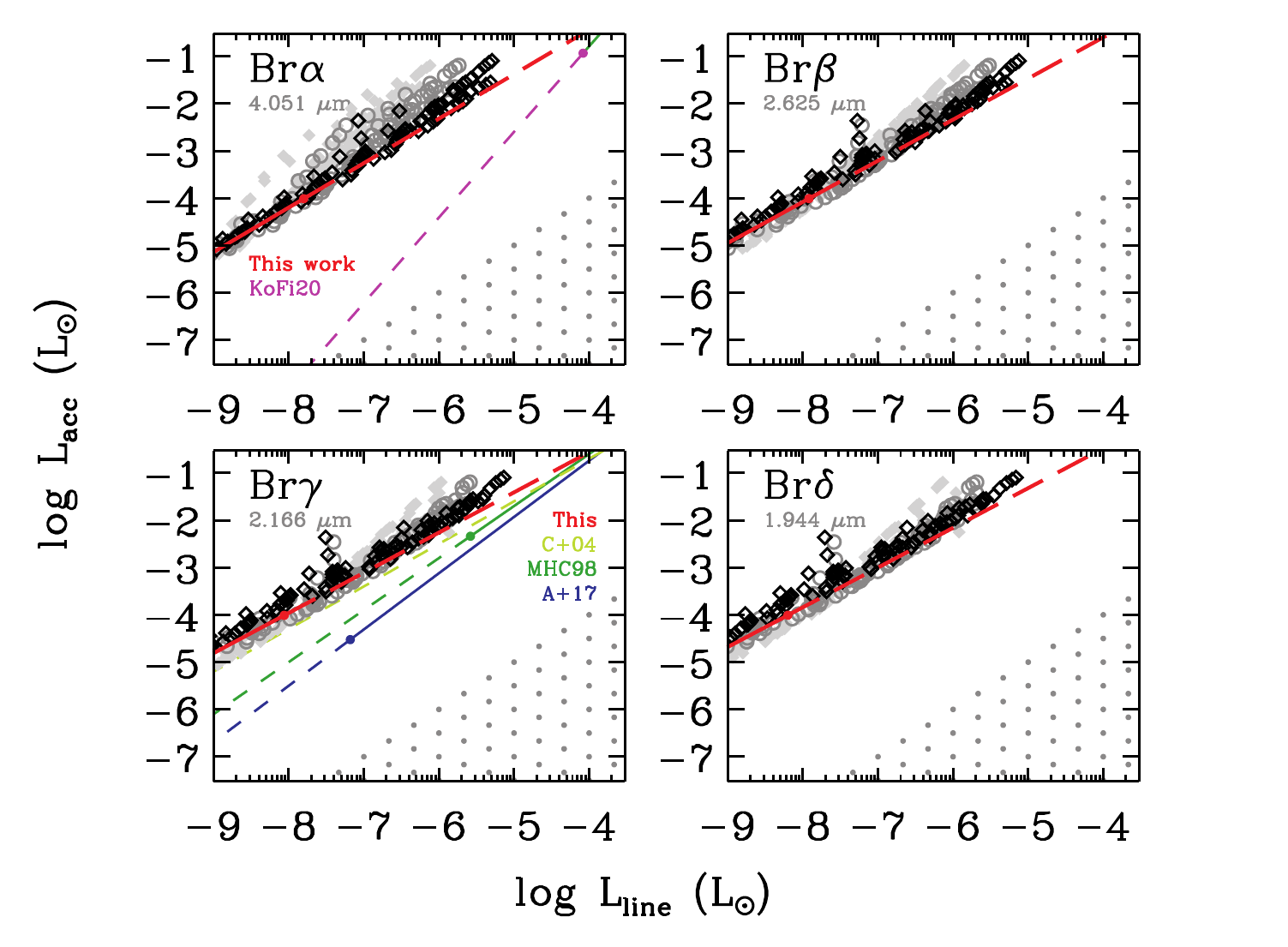}
    \caption{
As in Figure~\ref{fig:morelines1} but for the first Brackett-series lines.
The fit of \citet[][\textit{purple}]{Komarova+Fischer2020} is shown
for \Bra (observable with \textit{JWST}),
and those of
\citet[][\textit{lime}]{Calvet+2004} and
\citet[][\textit{green}]{Muzerolle+1998} for \Brg.
}
    \label{fig:morelines2}
\end{center}
\end{figure*}

\begin{figure*}
\begin{center}
\includegraphics[width=0.8\textwidth]{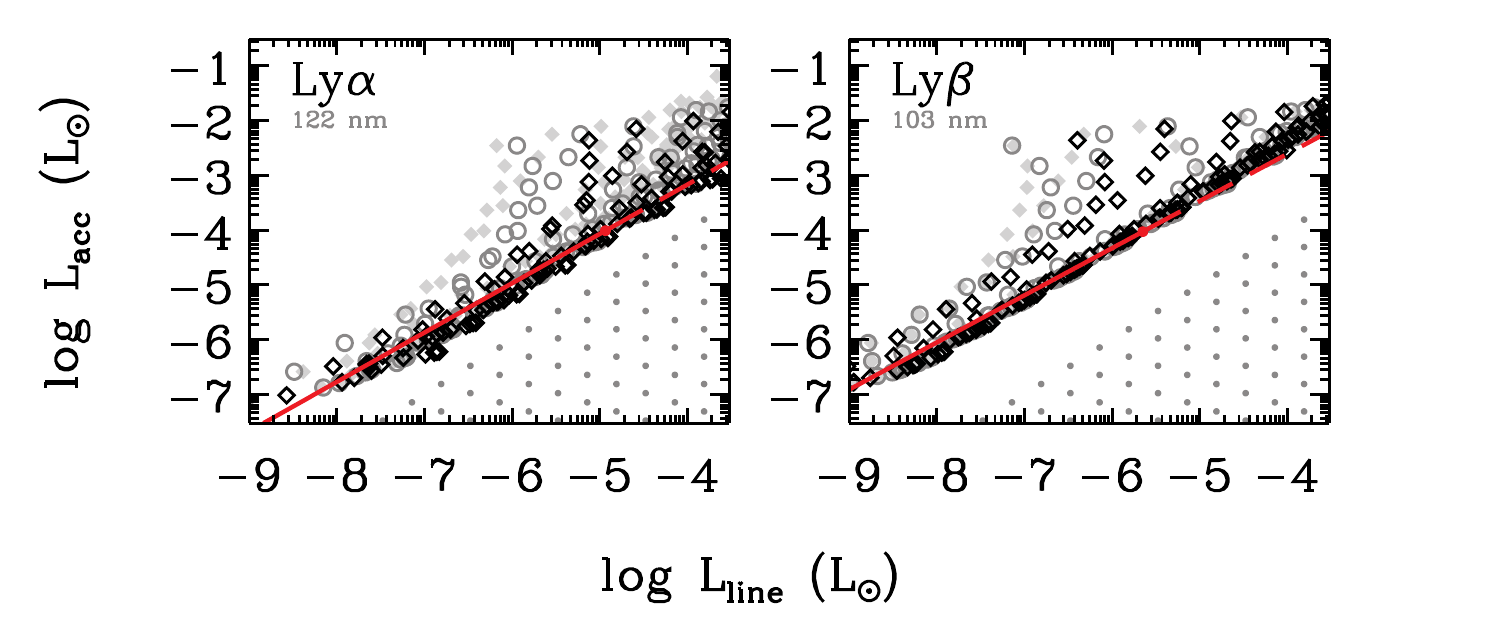}
\caption{
As in Figure~\ref{fig:morelines1} but for \Lya and \Lyb. Lyman-series lines are less reliable than the others but are shown for completeness.
}
    \label{fig:morelines3}
\end{center}
\end{figure*}

Figures~\ref{fig:morelines1}--\ref{fig:morelines3}  %
show our model results and the fits for several lines from the Balmer (\Ha,
\Hb, \Hg, \Hd),
Paschen (\Paa, \Pab, \Pag, \Pad),  %
and Brackett (\Bra, \Brb, \Brg, \Brd) series,
and also from the Lyman series (\Lya, \Lyb).  %
In all cases, the fit to our results (\textit{red line}) is roughly a lower limit.
For the chosen range of input planet masses (2--20~$\MJ$) and excluding the Lyman-series lines, the half-spread in \Lacc at a given \Lline is often relatively small, with $\sigma\approx0.3$~dex, but can reach $\sigma\approx0.5$~dex.
For the Lyman lines, the high optical depth of the upper layers of the postshock region lead to strong self-absorption. This is especially true for the points for lower planet masses, which have a higher density at a given \Mdot since $\rho\propto1/\vf$.

For all transitions, our data is above the stellar relationship, for the \Lline values covered both by their data and our model results, as well as where the stellar fit is extrapolated.
The difference reaches up to 1--2~dex for the Balmer lines, especially compared to \citet{Rigliaco+2012}, and 2--3~dex for Paschen lines. For \Bra, the difference is extreme (2--4~dex) compared to the fit of \citet{Komarova+Fischer2020}. This is however not surprising because there is barely an overlap in \Lacc between their data and our models, and their fit does not cover at all the \Lline values relevant to planetary accretion ($\Lline\lesssim10^{-5.5}~\Lsun$). In general,
as discussed in Section~\ref{sec:compLacc}, we do not expect the \Lacc--\Lline relationships to match between the stellar and planetary regimes because the generating mechanisms probably differ significantly. 
Note that, except for the \Ha fit of \citet{Rigliaco+2012}, none of the extrapolations of the stellar fits reaches into the $\Lline>\Lacc$ region, which would be likely unphysical (Section~\ref{sec:compLacc}).

\bibliographystyle{yahapj}
\bibliography{reference}

\end{document}